\documentclass{PoS}

\usepackage{subfig}

\title{Data model issues in the Cherenkov Telescope Array  project}

\ShortTitle{CTA Data Model}

\author{ J.L. Contreras $^a$, \speaker{K. Satalecka} $^a$, K. Bernl\"ohr $^b$ , C. Boisson $^c$, J. Bregeon $^d$ , A. Bulgarelli $^e$ ,G. de Cesare $^e$, R. de los Reyes $^b$ , V. Fioretti $^e$ ,K. Kosack $^f$ , C. Lavalley $^d$, E. Lyard $^g$, R. Marx $^b$, J. Rico $^h$, M. Sanguillot $^d$, M. Servillat $^c$, R. Walter $^g$, J.E. Ward $^h$ and A. Zoli $^e$   for the CTA consortium \footnote{Full consortium author list at http://cta-observatory.org}.\\

\llap{$ˆa$} UCM, Madrid, Spain.\\
\llap{$ˆb$} MPIK, Heidelberg, Germany.\\
\llap{$ˆc$} LUTH, Paris, France.\\
\llap{$ˆd$} LUPM, Montpellier, France.\\
\llap{$ˆe$} INAF/IASF, Bologna, Italy.\\
\llap{$ˆf$} CEA, Saclay, France.\\
\llap{$ˆg$} ISDC, Versoix, Switzerland.\\
\llap{$ˆh$} IFAE, Barcelona, Spain.\\

E-mail:\email{satalk@gae.ucm.es}
}

\abstract{The planned Cherenkov Telescope Array (CTA), a future ground-based Very-High-Energy (VHE) gamma-ray observatory, will be the largest project of its kind. It aims to provide an order of magnitude increase in sensitivity compared to currently operating VHE experiments and open access to guest observers. These features, together with the thirty years lifetime planned for the installation,  impose severe constraints on the data model currently being developed for the project.  

In this contribution we analyze the challenges faced by the CTA data model development and  present the requirements imposed to face them. While the full data model is still not completed we show the organization of the work, status of the design, and an overview of the prototyping efforts carried out so far. We also show examples of specific aspects of the data model currently under development.
}

\FullConference{The 34th International Cosmic Ray Conference,\\
		30 July- 6 August, 2015\\
		The Hague, The Netherlands}

\begin{document}

\section{Introduction}

\begin{figure}[htb]
\centering
\includegraphics[width=.95\textwidth]{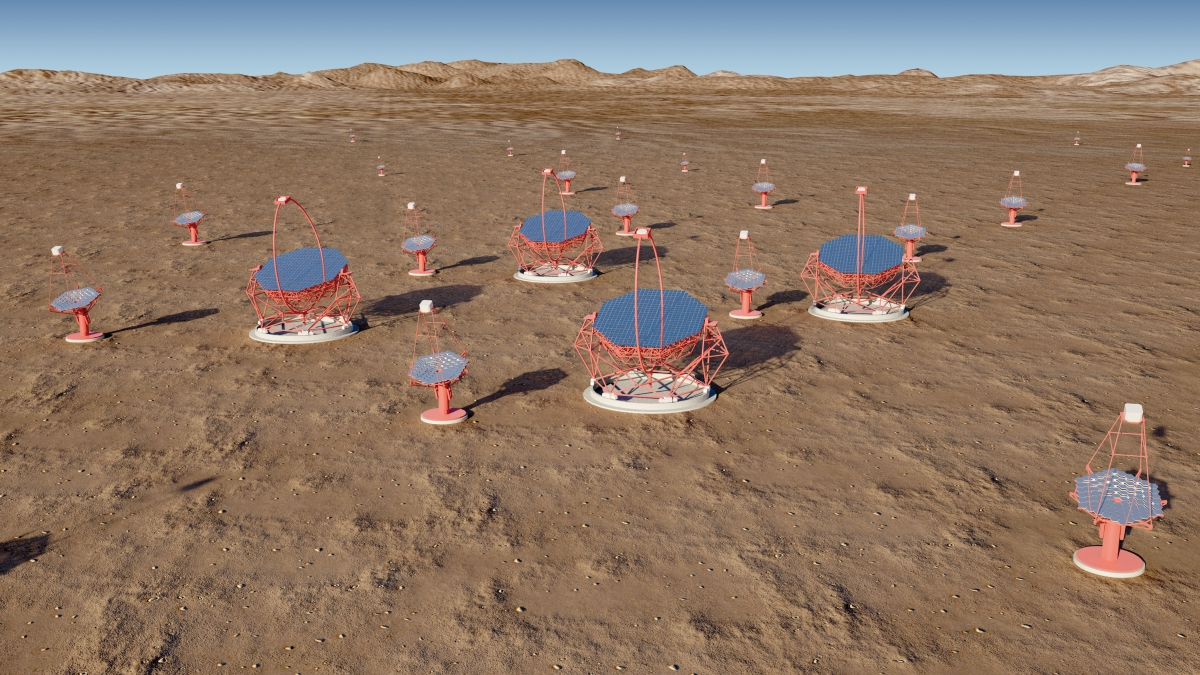}
\caption{Artistic view of the CTA Southern site (G. P\'erez IAC) }
\label{fig:CTA}
\end{figure}

  CTA \cite{cta-concept} can be considered as the first, ground based, Astroparticle Physics observatory. It aims at providing an order of magnitude increase in sensitivity compared to current VHE experiments and plans to operate for thirty years, covering the energy range between 20 GeV and 300 TeV. It will access the whole sky through two observatories located respectively  in the Southern and Northern hemispheres. The Southern observatory, with more complete access to the galactic plane, will deploy around 100 telescopes of three different types, to cover different energy ranges, while the Northern one will house around 20 telescopes of two different kinds. Several concepts have been developed for the CTA Telescopes and it is not yet decided whether all of them will be used in the final arrays. Still their common aspects are enough to allow them to be described by a common data model.
   
   Data will be processed on the observatory sites by both a real time and a delayed analysis chains to generate science alerts and monitor the instrument. Afterwards it will be transmitted to the off site data centers for the final analysis and storage \cite{cta-DM}. These centers will provide data access services to the scientists and technical personnel of the observatory.   
   
  Besides being able to efficiently cope with a large foreseen data rate, the CTA data management chain should provide open access to the data and implement stable formats that can last at least for the foreseen lifetime of the observatory, 30 years. In this paper we discuss how the CTA Data Model group is facing these challenges. We start in section \ref{data-rates} with an estimation of the data rates expected for CTA and follow by explaining the different types of data and data levels that have been identified along the data reduction chain. Section \ref{pbs} describes the structure of the products that will be provided by the group, which at the same time defines its present organization. The last two sections are devoted to the status of the design of the data model and some of the prototyping efforts carried out so far.

\section{Data rates}
\label{data-rates}

The gain in sensitivity required for CTA translates in a large foreseen trigger rate, more than two orders of magnitude higher than the one of present experiments such as H.E.S.S. or MAGIC. In each observatory four Large Size Telescopes with 23 m diameter dishes will reach rates around 10 kHz per telescope due to their low energy threshold. The Medium Size Telescopes, of 12 m dish diameter, aimed at intermediate energies will reach more moderate trigger rates, around 3 kHz, but their high number (24 in the South, 15 in the North) will more than compensate it in terms of data flow. Finally, in the South site, around 70 Small Size telescopes with rates around 400 Hz will cover the region of high energies and low fluxes. It is also planned to install innovative Schwarchild-Coud\'e Medium Size Telescopes in a later phase, only in the Southern observatory. Although their contribution to the data rates will likely be very significant we will not treat them in this note due to the remaining uncertainties about its value. All the numbers given above are derived from detailed Monte Carlo simulations of the arrays \cite{cta-mc}.

Cosmic Rays and VHE gamma-rays interact with the atmosphere giving rise to Extensive Air Showers that emit fast pulses of Cherenkov light. Imaging Atmospheric Cherenkov Telescopes (IACTs,) as those that will compose CTA, record images of the shower development. For this goal each type of telescope is equipped with a fast camera composed either of classical photo-multipliers (PMT) or silicon photomultipliers (SiPMs), in a number ranging from 1200 to more than 2000. Most of the cameras sample the light front for tens of nanoseconds, recording one sample per  nanosecond. The Cherenkov pulse will occupy a few nanoseconds inside the sampled window, but its position can only be known {\it a posteriori}.

There is a combination of four factors which leads to huge raw data rates: many telescopes, thousands of pixels per camera, trigger rates of some kHz and 30-100 samples per window. Around 300 Petabytes per year of operation would be produced by the arrays according to the Monte Carlo simulations, if all information is kept. As a work hypothesis at least a first step in data reduction has been assumed to take place before storage. It consists in keeping the whole set of samples only for a small set of pixels (3\% in average), those pertaining to the shower image, for the rest only an estimation of the total signal collected would be kept. The resulting data volumes, around 40 Petabytes, are still above the data volume that can be reasonably transported to the data centers and stored. Therefore the requirement to further reduce these data rates on site by a factor of 10 has been placed. It will be achieved by further suppressing empty pixels or events and applying data compression. An illustration of the basis of the data reduction procedure can be seen in figure\ref{fig:ICDB} (a). For pixels inside the ellipse the full waveform would be kept, while those outside would be integrated over time.

 The analysis pipelines of CTA will process the data from raw data acquired by the arrays to produce high level scientific products. They will also use as inputs Monte Carlo simulations and technical data acquired concurrently with the observations. The Data Model of CTA is based on defining several data levels along this chain. The lowest levels will be short lived, existing only in the  electronics reading the PMT signals or in buffers maintained by the Data Acquisition System.
 We define as data level 0, DL0, the set of data that will arrive to the CTA data centers and be stored there.
 Table \ref{tab:data-levels} resumes the data levels defined so far, not including short lived ones. The reduction factors have to be understood as indications and goals. Not all the data levels will be saved.

\begin{center}
\begin{table}[ht!]
    \begin{tabular}{| c | c | p{6cm} | c |}
    \hline
 Data Level & Short Name & Description & Reduction \\
 \hline
Level 0 (DL0) & DAQ-RAW & Data from the Data Acquisition hardware/software. & \\
 \hline
Level 1 (DL1) & CALIBRATED  & Physical quantities measured in each
                              separate camera: photons, arrival
                              times, etc.,  and per-telescope parameters derived
                              from those quantities.  & 1-0.2 \\
 \hline
Level 2 (DL2) & RECONSTRUCTED  & Reconstructed shower parameters (per
                                 event, no longer per-telescope) such
                                 as energy, direction,  particle ID,
                                 and related signal discrimination parameters. & $10^{-1}$ \\ 
\hline
Level 3 (DL3) & REDUCED  & Sets of selected (e.g. gamma-ray-candidate)
                           events, along with associated instrumental
                           response characterizations and any
                           technical data needed for science analysis.  & $10^{-2}$ \\
 \hline
Level 4 (DL4) & SCIENCE  & High Level binned data products like spectra, sky maps, or light curves.  & $10^{-3}$ \\
 \hline
Level 5 (DL5) &  OBSERVATORY  & Legacy observatory data, as survey sky maps or the CTA source catalog. & $10^{-5}$ - $10^{-3}$ \\
  \hline
    \end{tabular}
  \caption{ Data levels foreseen in CTA.}
\label{tab:data-levels} 
\end{table}
\end{center}

%Not all the data levels will be saved.

\section {Data Model products }
\label{pbs}
\begin{figure}[ht!b]
\centering
\includegraphics[width=1.0\textwidth]{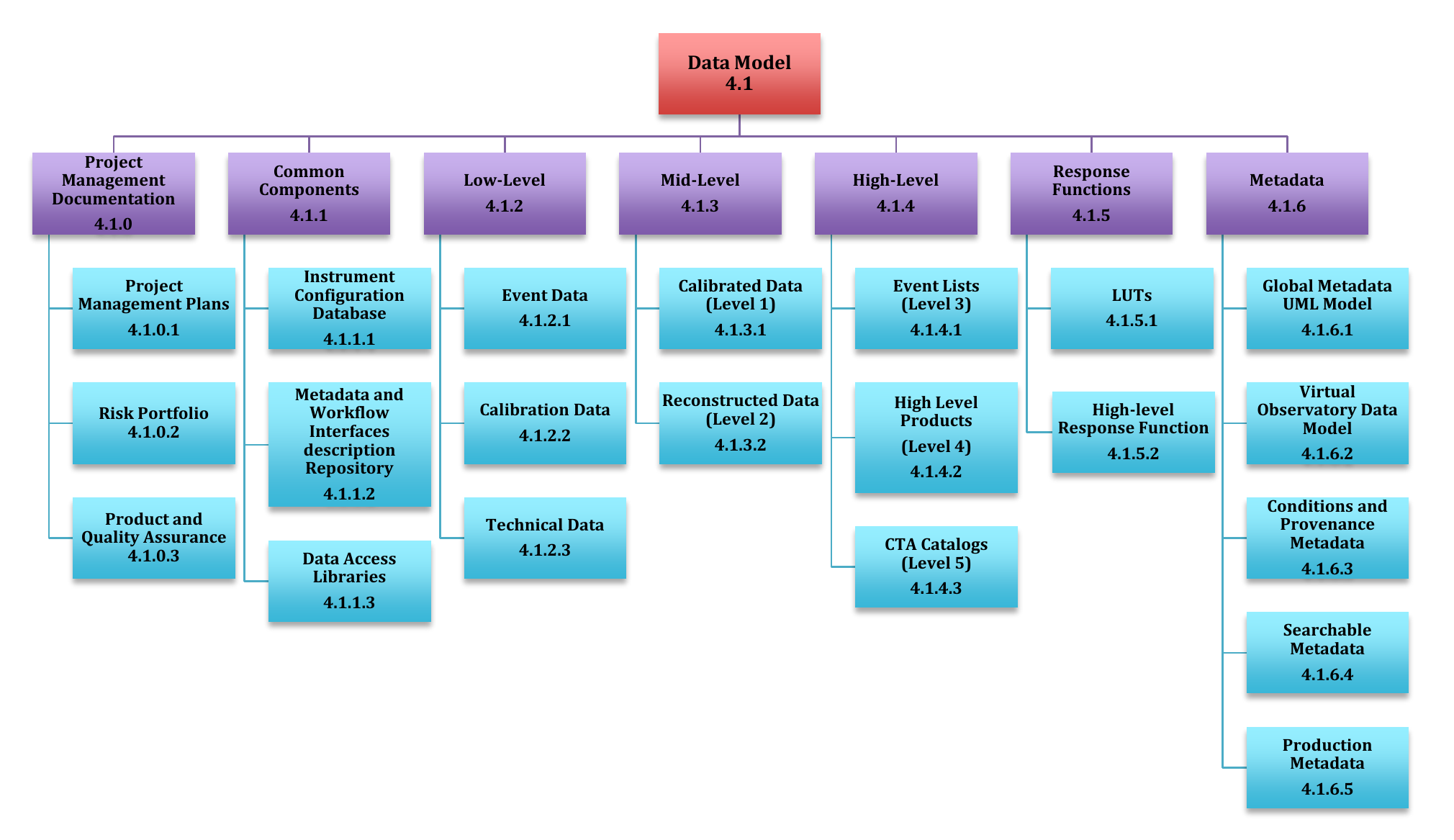}
\caption{Products for the CTA Data Model Workpackage}
\label{fig:DM-PBS}
\end{figure}

The first two levels of the Data Model working group Product Breakdown Structure (PBS) are presented in figure \ref{fig:DM-PBS}.
They all proceed from the Data Model Product, numbered as 4.1 in the CTA PBS. 
The PBS comprises six main products in addition to the work package documentation. The {\it Common Components} product (4.1.1) groups items that
are related to several data levels at the same time or whose aspects can affect several data levels. Three of them have been identified: the  {\it Instrument Configuration Database} (to keep the information related to array geometry, telescope geometry, etc), the {\it Data Access Libraries}, and the {\it Metadata and Workflow Interface Description Repository} which will contain the information about all
the Metadata and Data exchanged  among packages in the observatory.
The {\it Low, Mid and High Level data} products (4.1.2-4) group the definition of the Data Model for the data levels explained in the previous section. The model for  the Low Level data (DL0) is specially important since it is in interaction with the instrument and must absorb all of its complexity. To handle this complexity it has been subdivided in three different products: Event, Calibration and Technical data.  

The instrumental responses or Instrument Response Functions (IRFs) (4.1.5) describe the characteristics of
the instruments needed to extract the physical information. Examples of IRFs are the energy and angular resolution, or the effective detection area. Two different levels have been identified : Low level response functions, denoted as Look-Up-Tables (LUTs),
which are applied to reconstruct shower parameters (DL2 data), and High Level Instrument Response
Functions (HLIRF) used in the calculation of spectra and fluxes (DL4 data).
Finally, the role of the Metadata Product (4.1.6) is to define the set of metadata describing the data content. It is closely related to the task 
of easing the access to CTA data by the tools developed by the International Virtual Observatory Alliance (IVOA) collaboration. Nevertheless {\it Metadata}  must not only define the metadata related to IVOA, but also those concerning data provenance, or used for the production or discovery of and access to data. 

\section{Design and Status}
Despite a few choices remaining to be made the data model for CTA is almost complete.
The most advanced parts of data model are those which 
require a close collaboration with other groups and which will be needed in the nearest future.
There is a clear scheme for the Instrument  Configuration Database, itself part of the Common Components product.
In the definition of the  
DL0 several options have been proposed and are currently being tested. High level data and IRFs will be
provided to users through files using FITS formats. For DL3 data, composed of lists of events, and DL4 and DL5, the
development is being driven by interaction with the IVOA. For IRFs two competing formats are currently being tested.
While proposals and prototypes exist for intermediate data, their development will take place jointly with the one
of the pipelines which will use and produce them, since typically they will not be delivered openly.
In the next sections we briefly sketch the work being done.

\subsection{Common Components}
\begin{figure}%
\centering
\subfloat[Illustration of data reduction]{{\includegraphics[width=0.36\linewidth]{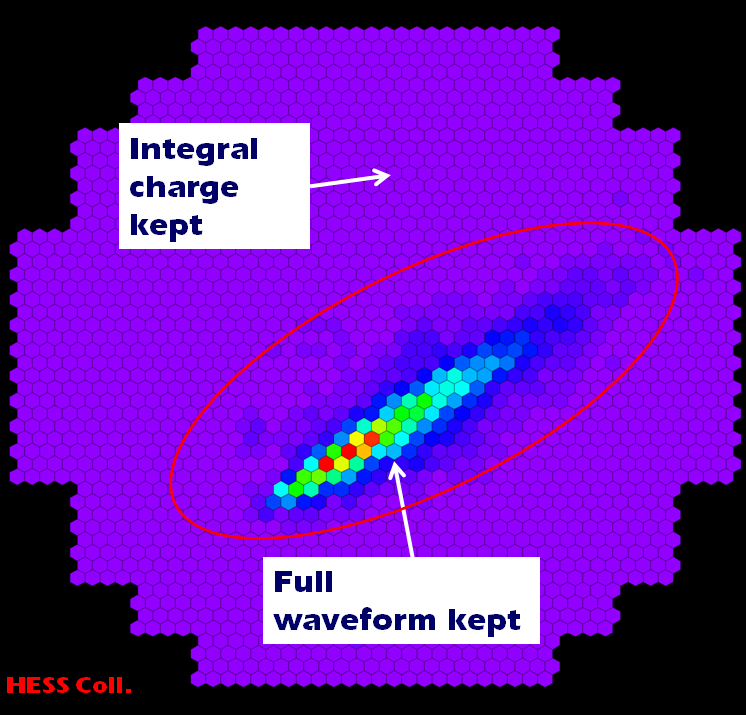}}}%
\qquad
\subfloat[ICDB Logical diagram.]{{\includegraphics[width=0.5\linewidth]{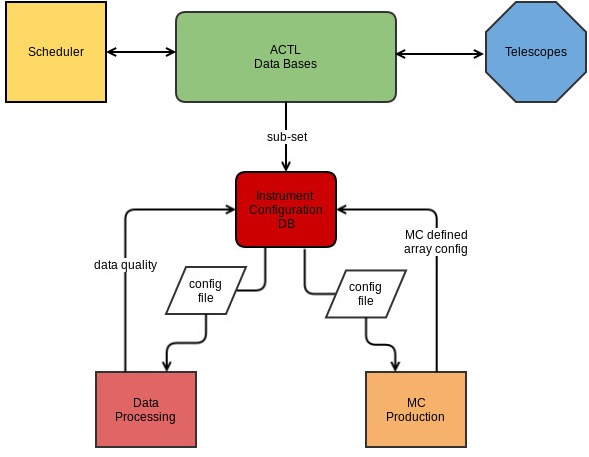}}}%
\caption{(a) Telescope image in the H.E.S.S. system of telescopes, explaining a possible data reduction scenario. (b) Logical diagram of the ICDB }
\label{fig:ICDB}
\end{figure}

Among the three products grouped as Common Components we single out the Instrument Configuration Database.
It is conceived as a repository, which can be thought as a database, to keep
information needed to define the instrument, e.g.: array coordinates, telescope types and positions,
camera types, etc.  It aims to reduce the dependence of the software on the time evolution of
the hardware. Along the life of the observatory some components will change more often than others, therefore
the database will have to be updated regularly. It will also contain Monte Carlo configurations, since
the simulations will likely use simplified or averaged descriptions of the instrument.
While the repository could be derived from similar products needed by other packages inside CTA, the
interface to the pipeline software will need to be coded ex-novo. Figure \ref{fig:ICDB}(b) shows a logical diagram of the
system.

\subsection{Low Level Data}
The Low Level Data, collectively called DL0, are defined as the lowest level of Event (EVT), Calibration (CAL) 
and Technical (TECH) data that are permanently archived. They come directly from the DAQ and might need to,
or have already been, modified on-site by some level of processing such as compression or zero suppression to
meet storage requirements. Its volume is determined by the amount of data produced by the cameras, EVT data, with
a contribution around 10-20\% from CAL and TECH.

The Data Flow for Low level data assumes that the images (EVT) from each telescope will be kept in separate files
together with some calibration and technical information needed for their first processing. EVTs from different
telescopes will only be merged once they are calibrated, at DL1, or in a preliminary process in the online  analysis.
This scheme eases the parallel processing. Each file will contain a time ordered chain of images from the camera of the telescope, acquired at the high rates imposed by the trigger and parallel chains of CAL and TECH information
acquired at lower frequencies.

The detailed content of DL0 data is presently being established in Interface Control Documents (ICDs) between the Data Management and Array Control groups and the groups building the cameras. Its main component will be the camera events, composed of camera information and different levels of pixel information depending on the data reduction level applied to each pixel.

For the format of the data and the files containing DL0 data three options are being considered and prototyped right now. One
of them is based on google protobuffs protocol~\cite{protobuff}  and compressed FITS~\cite{cfits}, another one in the
PACKETLIB~\cite{packetlib} format used  in space missions and a third one is an extension of the format presently 
used for the Monte Carlo data by the H.E.S.S and CTA collaborations~\cite{actio}.

\subsection{Intermediate Level Data}
Data of htis level will only be used by CTA pipelines and possibly internal CTA observatory staff.
Therefore their definition is more
open and will develop in conjunction with the pipeline work. A possibility which has been considered 
and tested is the use of the HDF5 file format.

Another option proposed is the use of Regions Of Interest (ROI), keeping only sections of the camera
surrounding the images. They allow to efficiently reduce the information, conserving a small fractions 
of pixels with no signal for calibration purposes. 
More information about a framework and file format based on the ROI approach, MESS, can be found in the contribution
published in these proceedings\cite{icrc-mess}. 

\subsection{High Level Data}
High Level data comprises the data levels that the CTA observatory will provide to guest observers and the scientific
community in general: DL3, DL4 and DL5. They must be provided in open, self documented formats. The observatory 
requirement is to use the FITS format.
 
Among the high level components the most important one is the DL3, consisting of lists of selected events
(eg. gamma rays or electrons) and the associated HL-IRFs needed to interpret them. DL3 data will be delivered
to guest observers together with a science tools 
package enabling them to tailor the analysis to their needs. Details on the observer access design for CTA can be found elsewhere in these proceedings\cite{icrc-oa}. A DL3 event will contain three kinds of information:
the quantities characterizing the particle (energy, direction, gamma/hadron tagging, etc.), those allowing to estimate
errors or retrieve the IRF information (uncertainties, number of telescopes used in the reconstruction, etc), and bookkeeping information (time, event number, etc.). A FITS format  has been defined following these lines, tested in a data challenge
and is being refined.

\subsection{IRFs} 

The response of the CTA arrays will depend on many correlated variables: characteristics of the primary particle (energy,   nature, incidence angle, etc), details of the detection process (number of telescopes implied, impact parameter, etc) 
atmospheric conditions etc. An optimal extraction of the physical quantities needs to take all of these parameters
into account in the IRFs, making their volume very large. Special data and file formats are being developed to cope 
with this problem. More information about one of the two considered approaches and its present status can be found in 
the dedicated contribution published in these proceedings\cite{icrc-irf}.

\subsection{Metadata}
The Metadata group is working towards defining a full set of metadata for CTA. The work has started by sketching the
global UML diagrams and then refining the description for different data levels.  
There is a close contact with the activities related to the IVOA.  
To ensure the integration of CTA data within the IVOA infrastructure the first step was to identify the
building blocks from existing IVOA data models suitable for description of gamma-ray data. This type of
data has never before been made publicly available in a common, open format. Current astronomical
metadata standards and VHE gamma-ray data conventions are being studied for this purpose, working
together with IVOA scientists. 

\section{Conclusions}
The design of the CTA Data Model is in an advanced status. It is based on the experience gained from previous
Cherenkov experiments plus the need to comply with the new requirements of open access, coping with
unprecedented data volumes and assuring long term stability.
A general scheme is already in place with advanced prototyping work existing for many of the components. 

\section*{Acknowledgments}
We gratefully acknowledge support from the funding agencies and organizations listed at the following
URL: http://www.cta-observatory.org


\begin{thebibliography}{999999}

\bibitem{cta-concept}
B. Acharya et al, CTA Consortium.
\emph{Introducing the CTA concept},
\emph{Astrop. Phys.} {\bf 43} (2013) 3-18.

\bibitem{cta-DM}
G. Lamanna et al, CTA Consortium.
\emph{Cherenkov Telescope Array Data Management},
\emph{in these proceedings}

\bibitem{cta-mc}
G. Maier et al,CTA Consortium.
\emph{Monte Carlo Performance Studies of candidate sites for CTA},
\emph{in these proceedings}


\bibitem{icrc-oa}
J. Kn\"odlseder et al, CTA Consortium.
\emph{Observer Access to CTA},
\emph{in these proceedings}

\bibitem{icrc-irf}
J.E. Ward, J Rico, T. Hassan, CTA Consortium.
\emph{The Instrument Response Function Format for the Cherenkov Telescope Array},
\emph{in these proceedings}

\bibitem{icrc-mess}
R. Marx and R. de los Reyes, CTA Consortium.
\emph{MESS: A Prototype for the Cherenkov Telescope Array Pipelines Framework}.
\emph{in these proceedings}

\bibitem{protobuff}
\emph{https://developers.google.com/protocol-buffers/}

\bibitem{packetlib}
A. Bulgarelli, F. Gianotti, M. Trifoglio.
\emph{PacketLib:  A C++ Library for Scientiﬁc Satellite Telemetry Applications}
\emph{ADASS XII, ASP Conference Series, Vol. 295, 2003}
\emph{https://github.com/ASTRO-BO/PacketLib}

\bibitem{actio}
K. Bernl\"ohr.
\emph{Simulation of imaging atmospheric Cherenkov telescopes with CORSIKA and sim\_telarray}
\emph{Astroparticle Physics, Vol 30,3, pp 149-158 (2008).}

\bibitem{cfits}
R.L. White et al.
\emph{Tiled Image Convention for Storing Compressed Images in FITS Binary Tables}
\emph{eprint arXiv:1201.1336}
\emph{http://fits.gsfc.nasa.gov/registry/tilecompression.html}

\end{thebibliography}
\end{document}